\documentclass[aps,prd,twocolumn,showpacs,showkeys,preprintnumbers,footinbib,floatfix]{revtex4}

\usepackage{graphicx}
\usepackage{amsmath, amsfonts, amssymb, bm}
\usepackage{epsfig} 
\usepackage{graphicx}
\usepackage{dcolumn}
\usepackage{xcolor}
\usepackage{longtable}
\usepackage{slashed}
\usepackage{enumerate} 
\usepackage[english]{babel}
\usepackage{textcomp}
\usepackage{hyperref}   
\usepackage{float} 
\usepackage{multirow}

\begin{document}

\title{Phenomenological model of multiphoto-production of charged~pion pairs on the proton}

\author{Anis Dadi}
\author{Carsten M\"uller}
\affiliation{Max Planck Institute for Nuclear Physics, Saupfercheckweg 1, 69117 Heidelberg, Germany}

\date{\today}

\begin{abstract}
The production of charged pion pairs via multiphoton absorption from an intense X-ray laser wave colliding with an ultrarelativistic proton beam is studied. Our calculations include the contributions from both the electromagnetic and hadronic interactions where the latter are described approximately by a phenomenological Yukawa potential. Order-of-magnitude estimates for $\pi^+\pi^-$ production on the proton by two- and three-photon absorption from the high-frequency laser field are obtained and compared with the corresponding rates for $\mu^+\mu^-$ pair creation.
\end{abstract}

\maketitle
%
%
\section{Introduction}
\label{intro}

Photoproduction of pion pairs on the proton has been studied extensively both in theory and experiment since the 1960s. In recent years the interest in the process has been revived by improved experimental data which were obtained by using polarized tagged photon beams at the MAMI synchrotron (Mainz, Germany) \cite{Daphne}, the GRAAL facility (Grenoble, France) \cite{GRAAL}, and the Jefferson laboratory (Newport News, USA) \cite{Jefferson}. At GRAAL, the high-energy photon beam is produced by Compton backscattering of laser light on a relativistic electron beam. These studies allow insights into the internal structure and excitation spectrum of the proton. A particular focus lies on polarization asymmetry measurements which are sensitive to interference cross terms. Various theoretical models based on effective Lagrangians for the pion-nucleon interaction have been developed (see \cite{Oset} and references therein).

When protons or other particles interact directly with intense laser beams (rather than with single photons from synchrotron or Compton backscattering sources), multiphoton processes may arise involving the simultaneous absorption of more than one photon. Multiphoton $e^+e^-$ pair creation was observed in ultrarelativistic electron-laser collisions at SLAC \cite{SLAC} and theoreticians have studied related $e^+e^-$ pair creation processes by relativistic proton impact on intense laser beams \cite{Report,MVG, Kaminski,  Milstein, Kuchiev, Baur}. In this setup, the laser frequency and field strength are largely Doppler-enhanced in the projectile rest frame. The process of $e^+e^-$ pair creation induced by high-energy neutrino impact on an intense laser beam has also been examined theoretically \cite{Tinsley}.

Inspired by the sustained progress in laser technology, very recently theoreticians started to study laser-induced $\mu^+\mu^-$ pair creation in proton-laser collisions \cite{Kuchiev, Deneke}. The apparent gap between the laser photon energy and the $\sim 100$\,MeV energy scale of the process can in principle be bridged by combining upcoming X-ray laser sources ($\hbar\omega_{\rm lab}\sim 10$\,keV) with the ultrarelativistic proton beam at CERN-LHC ($\gamma\approx 7000$) \cite{LHC}. The Doppler-upshifted photon energies $\omega\approx 2\gamma\omega_{\rm lab}\sim 100$\,MeV lie in the desired range. Large-scale \cite{XFEL} as well as table-top \cite{tabletopXFEL} free-electron lasers (FELs) are currently being developed aiming at the generation of intense coherent X-ray pulses. Coherent X-rays are also envisaged via high-order harmonics from oscillating plasma surfaces \cite{surface} or atomic gas jets \cite{Klaiber}. Such compact and portable X-ray sources hold the potential to be operated in conjunction with the LHC proton beamline.

In this paper, we consider $\pi^+\pi^-$ pair creation by multiphoton absorption in ultrarelativistic proton-laser collisions, i.e. the reaction $p+n\omega\longrightarrow p+\pi^+ +\pi^-$ with the photon number $n>1$. To this end, we combine the well-established approach to multiphoton processes in quantum electrodynamics (QED) \cite{strongfieldQED} with a simple phenomenological model to describe the pion-nucleon interaction. In general terms, the present study may be considered a first step towards an extension of the theory of laser-dressed QED into the realm of hadronic physics. More specifically, our main goals are (i) to provide order-of-magnitude estimates for $\pi^+\pi^-$ multiphoto-production rates, demonstrating the observability of the process; (ii) to compare with the corresponding rates for $\mu^+\mu^-$ production and show that a range of laser frequencies exists where $\pi^+\pi^-$ production dominates over the (direct) production of muons; and (iii) to discuss prospects why detailed investigations of multiphoto-production of pions might be useful.

We note that highly energetic reactions can also be induced when high-power laser pulses interact with solid targets. In the resulting plasma wakefields, electrons are accelerated to relativistic energies and emit secondary bremstrahlung $\gamma$-rays. These have led to the observation of photonuclear reactions \cite{LaserNuclear} and efficient $e^+e^-$ pair creation through the Bethe-Heitler effect \cite{Chen}. The setup also offers prospects for Bethe-Heitler creation of muon pairs \cite{Kaempfer} and single pion photoproduction through the reaction $\gamma p\to \pi^+n$ via the $\Delta(1232)$ resonance \cite{Karsch}. We point out that all these processes rely on single photo-absorption and do not exhibit multiphoton character. Particle reactions such as $\mu^+\mu^-$ and $\pi^+\pi^-$ production were also considered in an $e^+e^-$ plasma coupled to a photon field \cite{muonPLB}. Single pion production in strong magnetic fields \cite{ADP} and in collisions of laser-accelerated protons with nuclei have been studied theoretically as well \cite{pion}.

In the following, natural units with $\hbar = c = \epsilon_0 = 1$ are used.
%
%
\section{QED description of multiphoton muon pair creation}
\label{muon_creation}

We briefly review the laser-dressed QED approach to muon pair creation in the combined fields of an electromagnetic wave and an atomic nucleus \cite{Report,MVG,Kaminski, Kuchiev, Deneke}. Within the external-field approximation of QED, the total Lagrangian of the problem reads
\begin{eqnarray}
\label{L_QED}
{\mathcal L}={\overline \Psi}[i\gamma_\mu (\partial^\mu + ieA^\mu_L + ieA^\mu_C)-m]\Psi\,,
\end{eqnarray}
where $A_L^\mu$ and $A_C^\mu$ are the four-potentials of the laser wave and the nucleus, respectively, and $m$ is the muon mass. In the spirit of the Furry picture, one may split the total Lagrangian ${\mathcal L}={\mathcal L}_V+{\mathcal L}_C$ into an unperturbed part ${\mathcal L}_V={\overline \Psi}(i\gamma_\mu D^\mu - m)\Psi$ and the remaining interaction ${\mathcal L}_C=e{\overline \Psi}\gamma_\mu A^\mu_C\Psi$. Here, $D^{\mu}=\partial^{\mu}+ieA_L^{\mu}$ denotes the covariant derivative with respect to the laser field. The field theory of ${\mathcal L}_V$ is solved exactly by the so-called Dirac-Volkov states $\Psi_{p_\pm, s_\pm}$ which include the interaction of the leptons with the plane-wave laser field up to all orders~\cite{Itzykson}. The leptons are characterized by their four-momenta $p_\pm^\mu$ and spin projections $s_\pm$ outside the laser field. 

The Dirac-Volkov solutions may be used as basis states in perturbative calculations with respect to the remaining interaction with the nuclear field. As a result, the leading-order $S$-matrix element for multiphoton muon pair creation on a proton at rest reads
\begin{eqnarray}
\label{Smuon}
S_{\mu^+\mu^-} = -i\int d^4x\, {\overline \Psi}_{p_-,s_-}(x) \gamma^0 \Psi_{p_+,s_+}(x) V_C(x)\,,
\end{eqnarray}
where
\begin{eqnarray}
\label{Coulomb}
V_C(x) = eA^0_C(x) = \dfrac{e^2}{4\pi |{\boldsymbol x}|}
\end{eqnarray}
denotes the Coulomb potential energy in this frame and $\gamma^0$ is a Dirac matrix.

To be specific, we assume a monochromatic laser wave of circular polarization. Choosing the propagation direction along the $z$ axis, the corresponding four-potential in the radiation gauge reads
\begin{equation}
\label{laser_pot}
A_L^{\mu}(x)= (0,{\boldsymbol A}_L)\,,
{\boldsymbol A}_L(x)=A_0[{\bf e}_x \cos(kx)+{\bf e}_y \sin(kx)]
\end{equation}
where $k^\mu=\omega(1,0,0,1)$ is the wave four-vector and $(kx)=\omega t-{\boldsymbol k}\cdot{\boldsymbol x}$ is the laser phase. Although the laser field is treated as a classical electromagnetic wave, photons arise from a mode expansion of the oscillatory parts in Eq.\,\eqref{Smuon} which contains multiphoton processes of arbitrary order \cite{Report,MVG, Kaminski, Kuchiev, Deneke}.

From the $S$-matrix, the $\mu^+\mu^-$ production rate is obtained as
\begin{eqnarray}
\label{R}
R_{\mu^+\mu^-} = \int \frac{d^3{\boldsymbol p}_+}{(2\pi)^3} \int \frac{d^3{\boldsymbol p}_-}{(2\pi)^3}\sum_{s_+,s_-} |S_{\mu^+\mu^-}|^2.
\end{eqnarray}
The cross section of the process may be obtained by dividing out the photon flux $j=\omega A_0^2/4\pi$. As an alternative to the $S$-matrix approach, the laser-dressed polarization operator can be employed to calculate total pair creation rates \cite{Milstein}.

A few additional remarks are in order here. (1) For a proper description of muon pair creation in the field of a heavy nucleus, the finite nuclear extension needs to be taken into account in general \cite{Deneke}. Here we restrict our consideration to protons as projectiles, which may be treated as pointlike [see Eq.\,\eqref{Coulomb}] to a good approximation. (2) Being interested in order-of-magnitude estimates for the production rates, the recoil suffered by the proton during the process is ignored for simplicity. (3) We will consider X-ray laser fields where the value of the Lorentz-invariant parameter $\xi\equiv eA_0/m$ is much smaller than unity. In this regime of laser-matter coupling, a process involving $n$ photons could, in principle, be calculated within $n$-th order of perturbation theory in the photon field. We find it more convenient, however, to work within the framework of laser-dressed QED.
%
%
%
%
\section{Effective hadronic model of multiphoton pion pair creation}
\label{pion_creation}
Various models of $\pi^+\pi^-$ photoproduction on the nucleon have been developed which are based on effective Lagrangians for the pion-nucleon interaction \cite{Oset}. It was found that a proper description of the process requires the inclusion of more than 40 Feynman graphs. They consist of one-point, two-point and three-point diagrams with respect to the number of hadronic vertices. In the present case of interest, where the pions are created by the absorption of more than one photon, the number of relevant Feynman diagrams increases tremendously, which renders a consideration on this level almost prohibitorily involved. However, order-of-magnitude estimates for $\pi^+\pi^-$ multiphoto-production rates may be obtained by applying the following simplified model.

We treat the pion as a scalar particle whose dynamics is governed by the covariant Lagrangian
\begin{eqnarray}
{\mathcal L}= \left(D^{\mu}\phi\right)^*\left(D^{\nu}\phi\right)\left[\delta_{\mu\nu}-\mathcal{K}_{\mu\nu}\right]-m_\pi^2\phi\phi^*+{\mathcal L}_{\rm had},
\label{ld}
\end{eqnarray}
where $\phi$ denotes the scalar pion field, $m_\pi$ the pion mass, and $\delta_{\mu\nu}$ the Kronecker symbol. The symmetric contribution
\begin{eqnarray}
\mathcal{K}_{\mu\nu}=-\dfrac{\lambda_e+\lambda_m}{m_\pi} A_0^2 k_\mu k_\nu
\end{eqnarray}
represents a measure for the strength of the induced electric and magnetic dipole moments of the pion by the external photon field \cite{Kruglov}. It characterizes, to some degree, their underlying structure from quarks. The experimental value $\lambda_e+\lambda_m\approx0.16\times 10^{-4}~\mathrm{fm^3}$ of the sum of the electric and magnetic pion polarizabilities  was recently measured in Primakoff scattering of high-energy pions at CERN-SPS~\cite{ref:Abbon}.

The electromagnetic coupling of the pion to the laser field is contained in the unperturbed part, ${\mathcal L}_0\equiv{\mathcal L}-{\mathcal L}_{\rm had}$, of the Lagrangian in Eq.~\eqref{ld}. Similarly as in Sec.II, the field theory of ${\mathcal L}_0$ can be solved exactly. The corresponding scalar states $\phi_p$ were derived in \cite{Kruglov}. They represent generalizations  to composite particles of the usual scalar Volkov solutions of the Klein-Gordon equation in the presence of an electromagnetic wave. In fact, all terms in Eq.~\eqref{ld} multiplied by the Kronecker symbol contribute to the usual Lagrangian of a pointlike spinless boson in an external laser field, which is solved by these Gordon-Volkov states \cite{Report}.

The term ${\mathcal L}_{\rm had}$ in Eq.~\eqref{ld} describes the hadronic interaction between the pion and the proton. For the sake of simplicity of our model, we assume a scalar Yukawa coupling of the form
\begin{eqnarray}
\label{g}
{\mathcal L}_{\rm had} = g{\overline\psi}\psi\phi\,,
\end{eqnarray}
where $\psi$ denotes the nucleon field. We consider the coupling constant $g$ as a free parameter of our model which will be adjusted to experimental data. Note that in reality the pion-nucleon coupling is more involved and consists of pseudoscalar and pseudovector interactions of Yukawa type \cite{Walecka}. Nevertheless, the simple form in Eq.~\eqref{g} already incorporates main features of the hadronic interaction.

The leading-order $S$-matrix element for $\pi^+\pi^-$ multiphoto-production on the proton in the presence of an intense laser field may be written as
\begin{eqnarray}
\label{Spion}
S_{\pi^+\pi^-} = -i\int d^4x\, \left(\phi_{p_-}^*\partial_t\phi_{p_+}-\phi_{p_+}\partial_t\phi_{p_-}^*\right) V_{\rm had}(\vec{x})
\end{eqnarray}
with the Yukawa potential
\begin{eqnarray}
\label{Yukawa}
V_{\rm had}(x)=\frac{g^2}{4\pi}
\frac{{\rm e}^{-m_\pi |{\boldsymbol x}|}}{|{\boldsymbol x}|},
\end{eqnarray}
arising from the interaction \eqref{g}. Formally, this potential is associated with a scattering-type Feynman diagram of second order in the hadronic coupling constant. In the spirit of a meson-exchange model, it reflects the short-range nature of the hadronic interaction which represents an essential difference with the long-range Coulomb potential \eqref{Coulomb}. The total $\pi^+\pi^-$ production rate, $R$, is obtained from the $S$-matrix as in Eq.\,\eqref{R}, with the spin sum omitted.

\begin{figure}[h]
\centering
\includegraphics*[width=20pc]{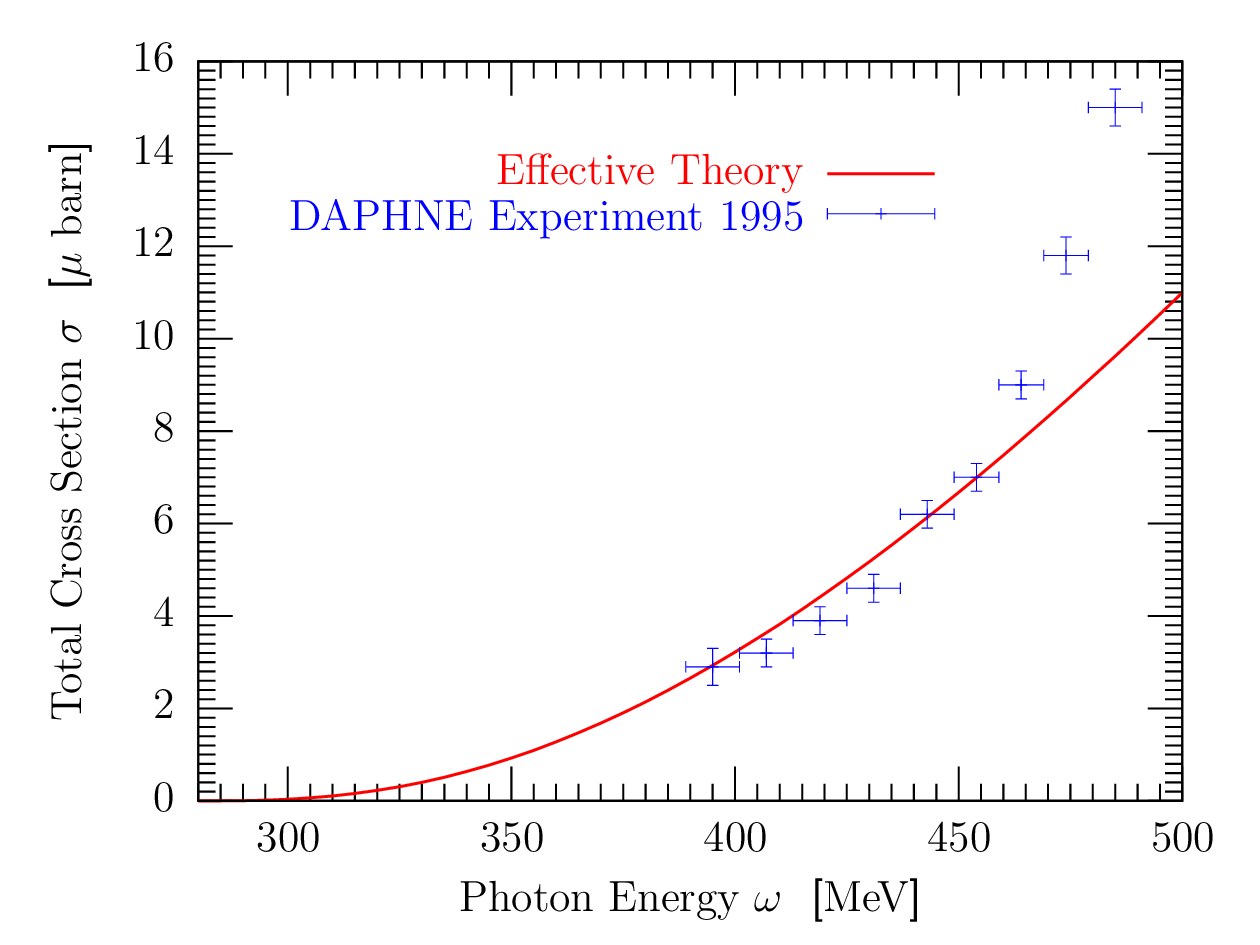}
\caption{(Color online) Total cross sections for $\pi^+\pi^-$ production by single-photon absorption on the proton. Shown are the experimental data recorded with the DAPHNE detector at MAMI \cite{Daphne} and the results from the present phenomenological model. The coupling constant for the pion-nucleon interaction was set to $g=4.3$ in order to reproduce the measured cross section at 450 MeV.}
\label{TC}
\end{figure}

We determine the numerical value of the coupling constant $g$ appearing in Eq.~\eqref{Yukawa} by taking reference to the available experimental data on $\pi^+\pi^-$ production by single-photon absorption on the proton. By setting $g=4.3$, we reproduce the $\pi^+\pi^-$ photoproduction cross section $\sigma = R/j$ at 450 MeV as measured at MAMI \cite{Daphne}. With this choice, we find reasonable agreement with the experimental data in the energy range from threshold up to $\approx 480$\,MeV (see Fig.~\ref{TC}).

Due to the adjustment of $g$ to a measured cross section, the Yukawa potential in Eq.~\eqref{Yukawa} becomes an effective potential which mimics the contributions from all relevant Feynman graphs in an approximative way. Our phenomenolgical model thus effectively accounts for various effects which are not included explicitely. These comprise, in particular, contributions from nucleonic resonances such as the $\Delta(1232)$.
Note that the value of the coupling constant extracted from measurements of pion-nucleon scattering is significantly larger, $g_{\pi N}\approx 13$ \cite{Arndt}, which clearly demonstrates the phenomenological character of our model.

At higher photon energies, our predictions nevertheless would deviate significantly from the experimental data in \cite{Daphne}. In fact, within our model the cross section raises approximately like a power law, $\sigma\sim (\omega-2m_\pi)^2$, and thus cannot reproduce the saturation occuring above 700 MeV observed in experiment. Such high energies, however, are not crucial in view of $\pi^+\pi^-$ production by multiphoton absorption. Namely, when a pion pair is created by the absorption of two photons with a total energy of 700 MeV, the energy of each single photon already lies above the $\pi^+\pi^-$ threshold. Since the probability for an $n$-photon process generally scales like $\xi^{2n}$ in the perturbative coupling regime ($\xi\ll 1$), the two-photon process is strongly suppressed as compared with the dominant single-photon process. Hence, for our purposes the good agreement in the energy range up to 480 MeV may be considered sufficient to obtain order-of-magnitude estimates for $\pi^+\pi^-$ multiphoto-production.

%
%
%
\section{Results on multiphoton $\pi^+\pi^-$ production}
Based on Eq.\,\eqref{Spion}, we have calculated total rates and angular spectra for multiphoto-production of $\pi^+\pi^-$ pairs in ultrarelativistic proton-laser collisions. The value $g=4.3$ was used for the pion-nucleon coupling constant in the effective Yukawa potential~\eqref{Yukawa}. The threshold for $\pi^+\pi^-$ production by two-photon absorption lies at an X-ray laser frequency of $\omega_{\rm lab}\approx m_\pi/2\gamma\approx 10$\,keV in the laboratory frame, when the proton beam is counterpropagating the laser beam with a Lorentz factor of $\gamma=7000$. This frequency domain is aspired by the currently emerging X-ray laser facilities \cite{XFEL,tabletopXFEL,surface,Klaiber}. Below, we mainly concentrate on processes involving two photons.

\begin{figure}[h]
\centering
\includegraphics*[width=20pc]{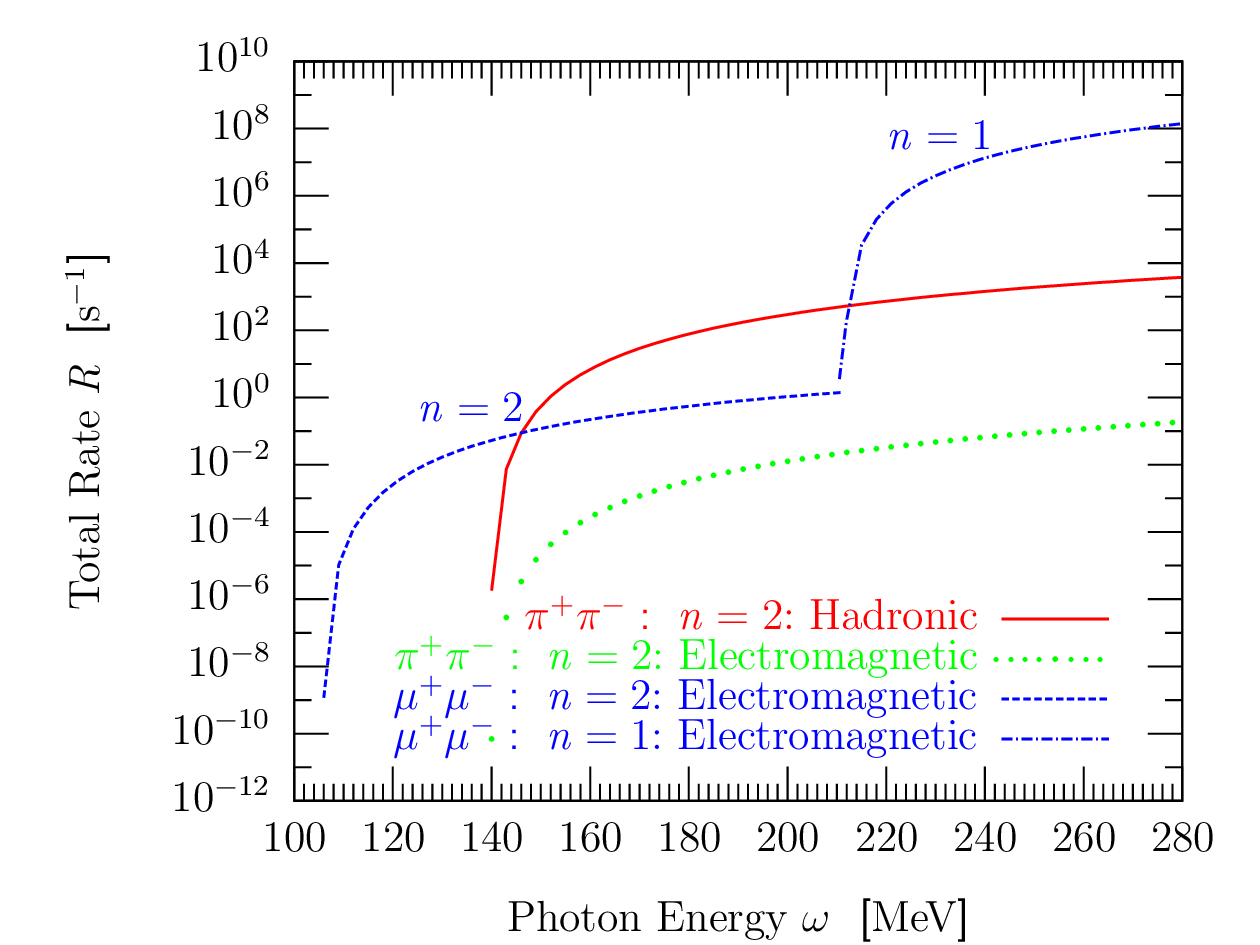}
\caption{(Color online) Total rates for $\pi^+\pi^-$ and $\mu^+\mu^-$ pair production on the proton by one- and two-photon absorption, as functions of the photon energy. The amplitude of the laser vector potential is $eA_0=5.1\,$keV, corresponding to lab-frame photon intensities $I_{\rm lab}=\omega_{\rm lab}^2 A_0^2/4\pi$ of the order of 10$^{22}$\,W/cm$^2$.}
\label{TR}
\end{figure}

Figure~\ref{TR} shows our results on the total production rates as a function of the photon energy, refering to the rest frame of the proton. In the laboratory frame, the rates are reduced to $R_{\rm lab}=R/\gamma$ due to time dilation.

Let us consider an example. For a photon energy of 200\,MeV in the proton frame, the $\pi^+\pi^-$ production rate by two-photon absorption amounts to $R\sim 10^3$\,s$^{-1}$ (see Fig.~\ref{TR}). Hence, in the collision of a proton beam containing $10^{11}$ particles \cite{LHC} with an X-ray pulse of 100\,fs duration \cite{XFEL}, the probability for production of one pion pair is $10^{-3}$, assuming perfect beam overlap.
An average $\pi^+\pi^-$ production rate of 10 events per second is obtained by taking the 
envisaged X-ray pulse repetition rate of 10\,kHz into account. Note that the latter can be synchronized with the revolution frequency of the LHC proton beam.

We note that the effect of the electric and magnetic pion polarizabilities turns out to be immaterial for the present parameters. The reason is that the parameter $(\lambda_e+\lambda_m)(p_\pm k)^2/m_\pi\sim (\lambda_e+\lambda_m)m_\pi^3\sim 10^{-5}$, which effectively quantifies the coupling of the external laser field to the pion polarizabilities \cite{Kruglov}, is much smaller than its coupling $e^2\approx 0.1$ to the pion charge.

In order to detect the process of $\pi^+\pi^-$ production by two-photon absorption unambigously in experiment, it is necessary to discriminate it from the sequential process where first a single $\pi^+$ and a neutron are created, $\gamma p \to \pi^+ n$, followed by the reaction $\gamma n \to \pi^- p$. For the purpose of discrimination, the angular distribution of the pions might serve as a fingerprint (see below).

It is interesting to compare the $\pi^+\pi^-$ production rate by two-photon absorption with the corresponding rate for $\mu^+\mu^-$ production \cite{Deneke}. Although the muon mass is substantially smaller than the pion mass, $\pi^+\pi^-$ production is dominant in the frequency range between $\approx 150$--210\,MeV, as Figure~\ref{TR} shows. For example, at a photon energy of 200\,MeV, the production of pions exceeds the production of muons by three orders of magnitude. The dominance of $\pi^+\pi^-$ production is due to the much larger strength of the hadronic interaction. The purely electromagnetic contribution to $\pi^+\pi^-$ production [resulting from the Coulomb potential of the proton in Eq.\,\eqref{Coulomb}] is suppressed by four orders of magnitude and, thus, smaller than the $\mu^+\mu^-$ rate. For higher photon energies ($\omega>2m$), the channel of $\mu^+\mu^-$ production by a single photon opens and represents the dominant process.

Our results imply that in the frequency interval $150\,{\rm MeV}\lesssim\omega\lesssim 210$\,MeV muons are predominantly produced indirectly via two-photon $\pi^+\pi^-$ production and subsequent pion decay, $\pi^+\to\mu^+ + \nu_\mu$ and $\pi^-\to\mu^- + {\bar \nu}_\mu$.

\begin{figure}[h]
\centering
\includegraphics*[width=20pc]{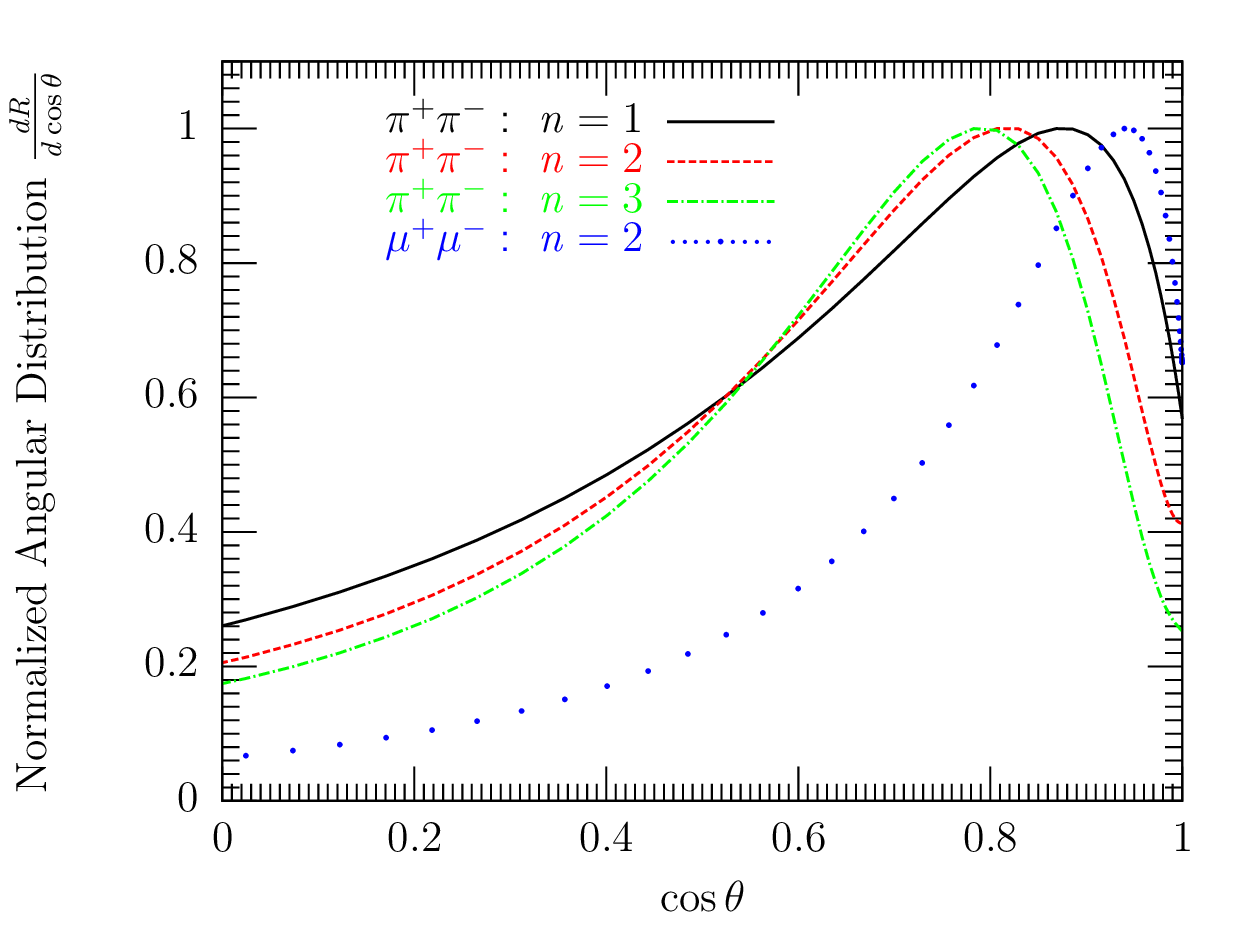}
\caption{(Color online) Angular distributions of one of the particles in $\pi^+\pi^-$ and $\mu^+\mu^-$ photoproduction on the proton. The spectra refer to the proton frame and the angle is measured with respect to the laser beam direction. In all cases, an identical amount of energy is absorbed from the laser field ($n\omega=360$\,MeV). The spectra have been normalized to the same height in order to facilitate their comparison.}
\label{DS}
\end{figure}

Figure~\ref{DS} displays our results on the angular distribution $dR/d\cos\theta$ of one of the created pions by one-, two- and three-photon absorption. The respective photon energies were chosen such that the energy absorbed in total is always 360\,MeV. The polar emission angle $\theta$ is measured with respect to the incident laser wave vector $\boldsymbol{k}$. 

The calculated angular spectrum for $\pi^+\pi^-$ production by one photon attains a maximum at $\theta\approx 29^{\circ}$. It is in reasonable agreement with experimental observation where a maximum around $\approx 30^{\circ}$ was found in the integrated energy interval from 350 to 500\,MeV \cite{spectrum}. For two- and three-photon absorption the position of the maximum is shifted towards larger angles ($\theta\approx 35^{\circ}$ and $\theta\approx 38^{\circ}$, respectively) and the distributions become slightly more narrow. These are characteristic multiphoton effects which have also been found in $e^+e^-$ pair creation by few-photon absorption \cite{Sarah}.

We draw again a comparison with two-photon production of $\mu^+\mu^-$ pairs. As Figure~\ref{DS} shows, here the angular distribution is peaked at smaller angles around $\theta_+\approx25^{\circ}$ and the contribution from large angles is suppressed. The differences are caused by the long-range nature of the Coulomb potential \eqref{Coulomb} as compared to the short-ranged Yukawa potential \eqref{Yukawa}. In the latter case, the particles are mainly produced at short distances which gives rise to large momentum transfers ${\boldsymbol q}$ corresponding to large emission angles. For the muons, small momentum transfers are relatively more important since the Fourier transform of the Coulomb potential is proportional to $1/{\boldsymbol q}^2$, as compared to $1/({\boldsymbol q}^2+m_\pi^2)$ for the Yukawa potential.

%
%
%
\section{Conclusion and Outlook}
A simple phenomenological model of $\pi^+\pi^-$ production by multiphoton absorption on the proton has been presented. The electromagnetic interaction of the pions with the photon field was incorporated into scalar Volkov states, while an effective Yukawa potential was used to describe the hadronic pion-nucleon coupling. The model contains one free parameter which has been adjusted to a measured cross section of $\pi^+\pi^-$ single-photon production at 450 MeV. Multiphoto-production of pion pairs could, in principle, be realized by utilizing the 7 TeV proton beam at CERN-LHC in conjunction with a compact 10--15\,keV X-ray laser source.

Predictions for total rates of two-photon $\pi^+\pi^-$ production were provided. In particular, it was shown that the two-photon production of $\pi^+\pi^-$ pairs substantially dominates over the (direct) production of muon pairs for photon energies above 150 MeV up to the threshold for $\mu^+\mu^-$ single-photoproduction. In this range of photon energies, muons are thus predominantly created indirectly via pion decays. We note that a similar conclusion was drawn in \cite{Kaempfer} with respect to pion and muon production by single bremsstrahlung photons in a laser-generated plasma.

Angular pion spectra have also been calculated which reflect the short-range nature of the hadronic interaction. When the $\pi^+\pi^-$ pair is created by more than one photon, larger emission angles arise which correspond to larger momentum transfers in the process.

As an outlook, we briefly discuss some general issues for which the process of multiphoto-production of pions might prove to be especially useful.

(1) Due to their long wavelength, photons interact uniformly over the nuclear volume and can therefore provide quantitative information on the entire structure of a given nucleus. Multiphoton processes such as $\pi^+\pi^-$ production might be particularly sensitive to the inner nuclear structure since they involve larger momentum transfers. The enhanced contribution from large momentum transfers moreover implies that multiphoto-production of pions probes a different spatial region inside the nucleus. This might help to gain an improved understanding of the short-range repulsion and intermediate-range attraction in the nucleon-nucleon potential.

(2) A main focus of pion photo-production is presently lying on polarization observables \cite{GRAAL,Jefferson}. Pion pair production via multiphoton absorption from a circularly polarized laser beam could be of particular interest in this regard. Since all photons carry the same helicity, the simultaneous absorption of several of them might lead to characteristic signatures on the polarization properties of the produced pions.

(3) Finally, photoproduction of pions exhibits an interesting analogy in atomic physics. A formal relation between the strong-field processes of $e^+e^-$ pair production and atomic photoionization in intense laser fields has already been revealed (see, e.g., \cite{MVG}). In the same spirit, a connection between photoionization of atoms and photoproduction of pions may be established, taking into account that pions can be produced both as single particles and in pairs. An analogy with single and double ionization of helium by one- and two-photon absorption is therefore suggestive. These four different ionization processes have found a sustained interest in atomic physics because of their sensitivity to electron-electron correlations \cite{helium}. While the production of single pions and pion pairs by one photon has been examined for a long time, we presented here a first study of pion pair production by two-photon absorption. The picture consisting of all four channels may be completed by a consideration of single pion production by two-photon absorption. 

We hope that our study may stimulate further research on multiphoto-production of pions on the nucleon.

\section*{Acknowledgements}
We thank B. Kopeliovich, H. J. Pirner, and B. Povh for helpful discussions on various aspects of hadronic interactions.

\end{document}